\documentclass[proceedings]{JHEP3} 
\PrHEP{ }

\usepackage{epsfig,multicol}        

\newbox\mybox
\newcommand\fverb{\setbox\mybox=\hbox\bgroup\verb}
\newcommand\fverbdo{\egroup\medskip\noindent\fbox{\unhbox\mybox}\ }
\newcommand\fverbit{\egroup\item[\fbox{\unhbox\mybox}]}

\title{Theoretical Predictions for the Direct Detection
of Supersymmetric Dark Matter}

\author{
{Carlos Mu\~noz} 
\thanks{This work was supported in part by the Spanish DGI of the
MCyT under Acci\'on Integrada Hispano-Alemana HA2002-0117,
and 
under contracts BFM2003-01266 and FPA2003-04597;
and the European Union under contract 
HPRN-CT-2000-00148. 
}\\
   Departamento de F\'{\i}sica Te\'orica C-XI and Instituto de 
F\'{\i}sica Te\'orica C-XVI,\\ 
Universidad Aut\'onoma de Madrid,
Cantoblanco, 28049 Madrid, Spain.\\
   E-mail: \email{carlos.munnoz@uam.es}
}

\conference{
International 
Workshop on 
Astroparticle and High Energy Physics
}

\abstract{
We compute the neutralino-nucleon cross section in several 
supersymmetric scenarios, taking
into account all kind of experimental and astrophysical constraints.
In addition, 
the constraints that the absence of dangerous charge
and colour breaking minima imposes on the parameter space
are also considered.
This computation is relevant for the theoretical 
analysis of the direct detection of
dark matter in current experiments.
We discuss interesting supergravity and superstring
scenarios.
}

\def\lsim{\raise0.3ex\hbox{$\;<$\kern-0.75em\raise-1.1ex\hbox{$\sim\;$}}}
\def\gsim{\raise0.3ex\hbox{$\;>$\kern-0.75em\raise-1.1ex\hbox{$\sim\;$}}}
\def\bmat{\left(\begin{array}}
\def\emat{\end{array}\right)}
\def    \be            {\begin{equation}}
\def    \ee            {\end{equation}}
\def    \bea           {\begin{eqnarray}}
\def    \eea           {\end{eqnarray}}

\begin{document} 

\section{Introduction}

\begin{figure}[t]
\begin{center}
\epsfig{file=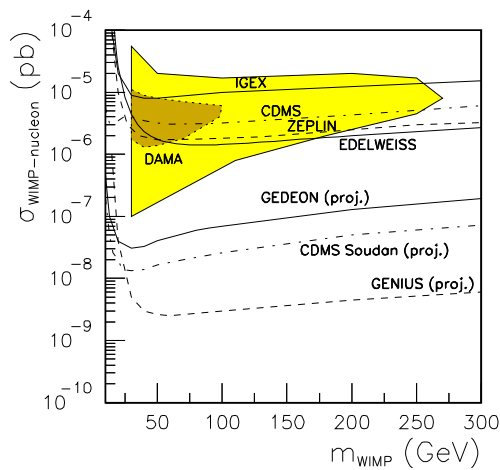,width=7cm}
\end{center}
\caption{
Areas allowed by the different experiments for the direct 
detection of dark matter in the parameter
space 
($\sigma_{\mbox{\tiny WIMP-nucleon}}$, $m_{\mbox{\tiny WIMP}}$).
The light shaded (yellow) area is allowed by the DAMA 
experiment, including uncertainties on the halo model.
The dark shaded (brown) area inside the previous one corresponds to the simple
case of an isothermal sphere halo model.
This case is also the one analyzed by the other experiments: 
The (lower) areas
bounded by solid, dot-dashed, dashed and (again) solid lines are
allowed by IGEX, CDMS, ZEPLIN I 
and EDELWEISS  
current experimental
limits.
The (upper) areas bounded by solid, dot-dashed and dashed lines 
will be analyzed by the 
projected GEDEON, CDMS Soudan and GENIUS experiments.
\label{limits}}
\end{figure}

Impressive experimental efforts have been carried out since 1987
for the direct detection of dark matter through elastic scattering with nuclei
in a detector \cite{lightreview}.
In fact, one of the experiments, the DAMA collaboration\footnote{A complete 
list of references
for this and the other experiments discussed below can be found
in Ref.~\cite{lightreview}.}, has
reported 
data favouring the existence of a 
Weakly-Interacting-Massive-Particle
(WIMP) signal. 
This signal is compatible 
with WIMP masses up to 100 GeV
and WIMP-nucleon cross sections in the interval
$10^{-6}-10^{-5}$ pb,
as shown with a dark shaded area in Fig.~\ref{limits}.

Notice, however, 
that this result has been obtained assuming the
simple isothermal sphere halo model with a dark-matter density
$\rho_0 = 0.3$ GeV\ cm$^{-3}$ 
and a 
Maxwell-Boltzmann local
velocity distribution for the WIMPs 
$f(v)\propto e^{-v^2/v_{0}^2}$,
with
$v_0= 220$ km\ s$^{-1}$.
When uncertainties on the halo model
are taking into account,
the signal is consistent with a larger region of the
parameter space.
In particular, in 
Ref.~\cite{halo} 
modifications in the velocity distribution function
for different galactic halo models were considered, using
in addition the allowed ranges for $v_0$ and $\rho_0$ in each model.
The final result of the analyses is shown in Fig.~\ref{limits} 
with a light shaded area. One sees that 
the signal is compatible with larger values of the parameters, 
i.e. WIMP masses up to 270 GeV
and WIMP-nucleon cross sections in the interval
$10^{-7}- 6\times 10^{-5}$ pb.
In fact, as discussed also in 
Ref.~\cite{halo}, 
when co-rotation of the galactic halo is also considered,
the mass range extends further to $500-900$ GeV
for cross sections in the interval
few$\times 10^{-6}- 2\times 10^{-5}$ pb.


Unlike this spectacular result, other collaborations 
claim to have excluded important regions of the DAMA 
parameter space, as shown also in Fig.~\ref{limits}.
In particular, the first of these was the CDMS experiment
A small part of the region excluded by CDMS
has also been excluded by IGEX 
and 
HDMS. 
But more disturbing 
are the recent results from 
EDELWEISS 
and 
ZEPLIN I 
collaborations,
excluding even larger regions than CDMS.
Let us remark that,
unlike DAMA,
for CDMS and for the other experiments analyses
taking into account the uncertainties in the galactic halo are not shown 
in the figure, and we only see the effect of the
standard halo model on their results. Including
those uncertainties, the light shaded area favoured by DAMA 
and not excluded
by the null searches would be in principle smaller than the one
shown here (for an analysis of this issue see Ref.~\cite{KKrauss}).

Owing to this controversy
between DAMA and the other experiments,
one cannot be sure whether or not the first direct evidence for the existence
of dark matter has already been observed.
Fortunately, the complete DAMA region will be tested in the future,
since
more than 20 experiments are running or in preparation around the
world.
This is for example the case of GEDEON, 
an expansion of
CDMS
called
CDMS Soudan, and 
GENIUS,
shown in Fig.~\ref{limits}.

Given this situation, and assuming that the dark matter 
is a WIMP, it is natural to wonder how big 
the cross section for its direct detection can be.
The answer to this 
question depends on the particular WIMP considered.
The leading candidate in this class is the lightest 
neutralino \cite{lightreview}, 
a particle 
predicted by the supersymmetric (SUSY) extension of the standard model.
It is then crucial to re-analyze the compatibility of the 
neutralino as a dark matter candidate, with the sensitivity
of present (and future) dark matter detectors.
Here we will concentrate on the most recent results in the context of supergravity
(SUGRA) and superstrings.
Concerning the latter, let us recall that
the neutralino is also a candidate for dark matter
in superstring constructions, since 
generically the low-energy limit of
superstring theory
is $4$-dimensional 
SUGRA.

Let us finally remark that we will impose in our computation
the usual experimental and astrophysical 
constraints in this type of analysis, namely
the lower bound on the Higgs mass,
the $b\to s\gamma$ branching ratio, the
muon $g_{\mu}-2$, and the observed 
dark matter density of the Universe.
Likewise,
the constraints that the absence of dangerous charge
and colour breaking minima imposes on the parameter space 
will also be taken into account \cite{darkufb}.


\section{SUGRA predictions}

In this section we will analyse the 
minimal-supersymmetric-standard-model (MSSM) scenario
in the framework of SUGRA. 
In this framework a large
number of free parameters are present in general. 
In order to have predictive power one usually assumes 
that the soft parameters 
are 
universal at $M_{GUT} \approx 2\times 10^{16}$ GeV.
This is the so-called minimal supergravity (mSUGRA) scenario,
where there are only four free parameters: 
$m$, $M$, $A$, and $\tan \beta$. In addition, the
sign of $\mu$ remains also undetermined since only
$\mu^2$ is determined by the minimization of the Higgs effective 
potential
\begin{equation}
\mu^2 = \frac{m_{H_d}^2 - m_{H_u}^2 \tan^2 \beta}{\tan^2 \beta -1 } - 
\frac{1}{2} M_Z^2\ .
\label{electroweak}
\end{equation} 
%
In any case, 
the general situation for 
the soft parameters in SUGRA is to have a
non-universal structure.
In the next subsections we will analyze all these possibilities.
In addition, we will 
also consider the possibility of relaxing the GUT scale,
and study the case of an intermediate scale.

\subsection{mSUGRA scenario with a GUT scale
}

As is well known, 
in the mSUGRA scenario with a GUT scale the lightest neutralino
$\tilde{\chi}^0_1$ is mainly bino
and, as a consequence, the predicted neutralino-proton cross section
$\sigma_{\tilde{\chi}_1^0-p}$
is well below the accessible experimental regions for low and moderate
values of $\tan\beta$ \cite{lightreview}.
Let us recall that
$\mu^2$ given by Eq.~(\ref{electroweak}),
for reasonable values of  $\tan\beta$,
can be approximated as
$\mu^2\approx -m_{H_u}^2-\frac{1}{2} M_Z^2$.
Thus it becomes 
large since $m_{H_u}^2$ evolves towards large and negative values
with the scale.
In particular, $|\mu|$ becomes 
much larger than
$M_1$ and $M_2$. As can be easily understood from the
neutralino mass matrix, 
the lightest neutralino will then be mainly gaugino, and in particular
bino, since
at low energy $M_1=\frac{5}{3}\tan^2\theta_W M_2\approx 0.5 M_2$.
Now, the scattering channels through Higgs exchange 
are not so important.
In addition,
the (tree-level)
mass of the CP-odd Higgs $A$,
$m^2_A=m_{H_d}^2+m_{H_u}^2+2\mu^2$,
will be large 
because $\mu^2$ is large.
Since the heaviest CP-even Higgs,
$H$, is almost degenerate in mass with this, 
$m_H$ will also be large
producing a further suppression in the scattering channels.

\begin{figure}[t]
\begin{center}
\epsfig{file=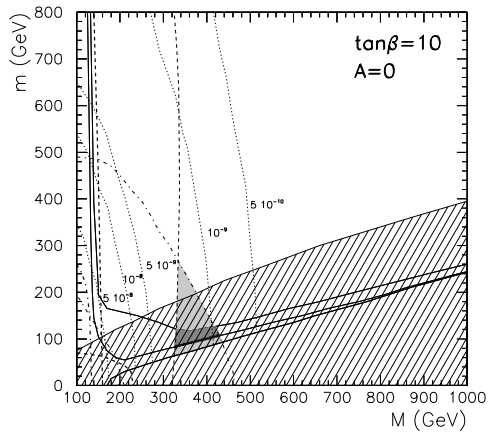,width=7cm}
\end{center}
\caption{Scalar neutralino-proton cross section $\sigma_{\tilde{\chi}_1^0-p}$
in the parameter space of the mSUGRA scenario ($m$, $M$) 
for $\tan \beta=10$, $A=0$
and $\mu > 0$.
The dotted curves are contours of $\sigma_{\tilde{\chi}_1^0-p}$.
The region to the left of the
near-vertical dashed line is excluded by the lower bound 
on the Higgs mass $m_h>114.1$ GeV. 
The region to the left of the near-vertical double dashed line
is excluded by the lower bound on the chargino mass
$m_{\tilde\chi_1^{\pm}}>103$ GeV.
The corner in the lower left shown also by a double dashed line
is excluded by the LEP bound on the stau mass
$m_{\tilde{\tau}_1}>87$ GeV.
The region bounded by dot-dashed lines is allowed by $g_{\mu}-2$.
The region to the left of the 
double dot-dashed line is excluded by $b\to s\gamma$.
From bottom to top, the solid lines are the upper bounds of the areas such
as $m_{\tilde{\tau}_1}<m_{\tilde{\chi}_1^0}$ (double solid), 
$\Omega_{\tilde{\chi}_1^0} h^2<0.1$ and $\Omega_{\tilde{\chi}_1^0} h^2<0.3$. 
The light shaded area is favored by all the phenomenological
constraints, 
while the dark one fulfills in addition
$0.1\leq \Omega_{\tilde{\chi}_1^0}h^2\leq 0.3$ (the black region on top of this
indicates the WMAP range $0.094<\Omega_{\tilde{\chi}_1^0}h^2<0.129$).
The ruled region 
is excluded because of
the charge and colour breaking constraint UFB-3.}
\label{a2m}
\end{figure}

This general fact is shown with an example in 
Fig.~\ref{a2m}, where contours 
of $\sigma_{\tilde{\chi}_1^0-p}$ in the
parameter space ($m$, $M$) for $\tan \beta=10$, $A=0$
and $\mu > 0$ 
are plotted\footnote{From now on
we will always show throughout the paper figures taken from Ref.~\cite{darkufb}
since they include the charge and colour breaking
constraints.}.
As we can see in the figure,
the experimental bounds mentioned in the Introduction
are very important and exclude large regions of the
parameter
space.
This is mainly due to the combination of the Higgs mass bound with
the $g_{\mu}-2$ lower bound\footnote{Note that we are using only the limit
based on $e^+e^-$ analysis,
$11.3\times 10^{-10}\leq a_{\mu} (SUGRA)\leq 56.1\times
10^{-10}$, otherwise the allowed region
would extend towards the right hand side of the figure.}. 
The light shaded area in the figure shows the
region allowed by the experimental bounds. There, the lower 
contour (double solid line)
is obtained including also
the constraint coming from the bound, 
$m_{\tilde{\chi}_1^0}<m_{\tilde{\tau}_1}$, 
in order to have an electrically neutral LSP.
For this area $\sigma_{\tilde{\chi}_1^0-p}\approx 10^{-9}$ pb.

On the other hand, when the astrophysical bounds
$0.1\lsim \Omega_{\tilde{\chi}_1^0}h^2\lsim 0.3$
are
also imposed the allowed area becomes very small. Only
the beginning of the tail where the LSP is almost degenerate with
the stau, producing efficient coannihilations, is rescued.
In addition, 
the restrictions coming from the charge and colour breaking
constraints,
in particular the UFB-3 one,
exclude also this area.

Let us recall that the UFB-3 direction, 
which involves
the scalar fields
$\{H_u,\nu_{L_i},e_{L_j},e_{R_j}\}$ with $i \neq j$
and thus leads also to electric charge
breaking, yields the strongest bound among all
the charge and colour breaking constraints. 
This is because 
the value of the potential along the UFB-3 direction is given
generically  
by
\bea
V_{\rm UFB-3}=(m_{H_u}^2
+ m_{L_i}^2 )|H_u|^2
+ \frac{|\mu|}{\lambda_{e_j}} ( m_{L_j}^2+m_{e_j}^2+m_{L_i}^2 ) |H_u|
-\frac{2m_{L_i}^4}{g^{\prime 2}+g_2^2} \ ,
\label{SU8}
\eea
and then it may become more negative than the MSSM potential at the minimum,
when $m_{H_u}^2$ evolves towards negative values.

In conclusion, the results indicate that the whole parameter space
for $\tan\beta=10$ is excluded on these grounds.
In fact, it is possible to show that 
$\tan\beta\lsim 20$ is excluded for any value of $A$ \cite{darkufb}.

\begin{figure}[t]
\epsfig{file=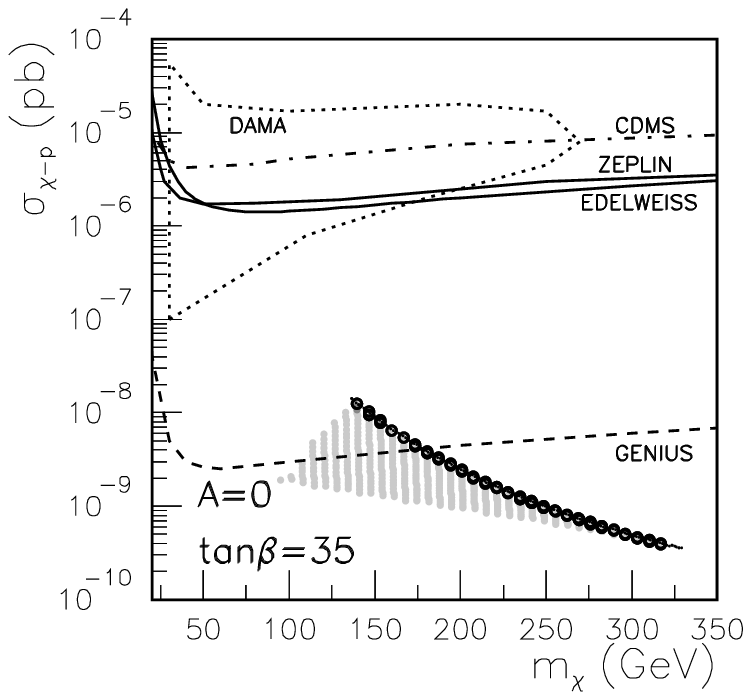,width=7cm}%
\epsfig{file=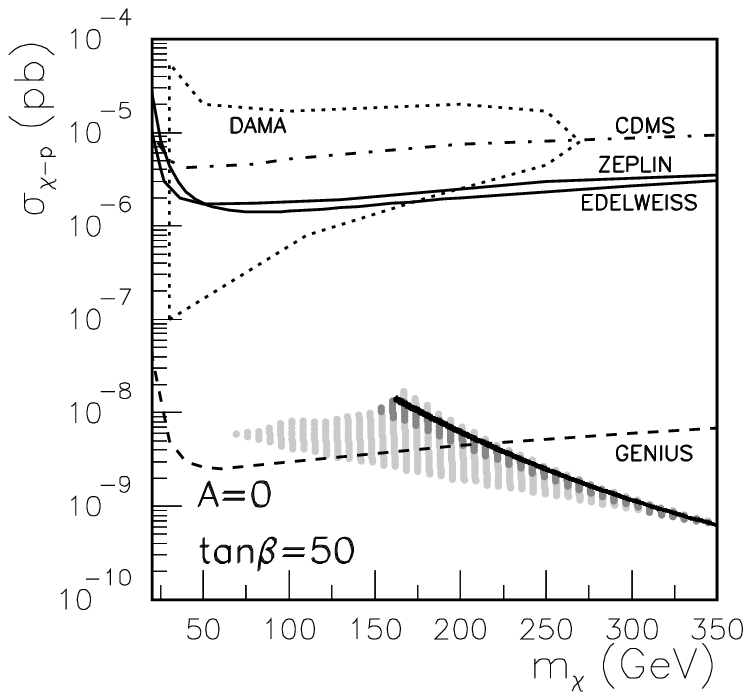,width=7cm}
\caption{Scatter plot of the scalar neutralino-proton cross section
\protect$\sigma_{\tilde{\chi}_1^0-p}$ 
as a function of the neutralino mass 
$m_{\tilde{\chi}_1^0}$ in the mSUGRA scenario,
for $\tan\beta=35$ and 50, $A=0$ and $\mu>0$. 
The light grey dots correspond to points
fulfilling all experimental constraints.
The dark grey dots correspond to points fulfilling in addition
$0.1\leq \Omega_{\tilde{\chi}_1^0}h^2\leq 0.3$
(the black dots on top of these
indicate those fulfilling the WMAP 
range $0.094<\Omega_{\tilde{\chi}_1^0}h^2<0.129$). 
The circles indicate regions 
excluded by the UFB-3 constraint.
The lines corresponding to the different experiments
are as in Fig. \protect\ref{limits}.}
\label{cross_scale122}
\end{figure}

\vspace{0.2cm}

The neutralino-proton cross section can be increased when the value
of $\tan\beta$ is increased 
\cite{Bottino}-\cite{large}.  
Notice for instance that the contribution of the
down-type quark to the cross section is proportional to
$1/\cos\beta$.
In addition, the bottom Yukawa
coupling increases, and as a consequence 
$m_{H_d}^2$ decreases, implying that
$m^2_A=m_{H_d}^2+m_{H_u}^2+2\mu^2$
also decreases. Since, as mentioned above, $m_H\approx m_A$,
this will also decrease significantly.
Indeed, scattering channels through Higgs exchange 
are more important now and their contributions to the cross section
will increase it.
Thus, in principle, we can even enter in the DAMA region.
However, 
the present experimental constraints 
exclude this possibility \cite{EllisOlive,Arnowitt3}.
We summarize this result
in Fig.~\ref{cross_scale122}. 
There, the values of $\sigma_{\tilde{\chi}_{1}^{0}-p}$
allowed by all experimental constraints, 
as a function of the neutralino mass
$m_{\tilde{\chi}_1^0}$, are shown
for
$\tan\beta=35,50$ and $A=0$.
Dark grey dots 
correspond to 
those points having a relic neutralino density within
the preferred range $0.1\leq\Omega h^2\leq 0.3$.
Given the narrow range of these points for the case 
$\tan\beta=35$,
they overlap 
in 
the figure with those
excluded by the UFB-3 constraint (shown with circles).
We observe that, generically, the cross section and the neutralino mass
are constrained to be (for any value of $A$)
$5\times 10^{-10}\lsim \sigma_{\tilde{\chi}_1^0-p}\lsim 3\times 10^{-8}$ pb and
$120\lsim m_{\tilde{\chi}_1^0}\lsim 320$ GeV, respectively.

\vspace{0.2cm}

It is possible to show that
qualitatively similar results are obtained when dark matter
is analyzed
in the so-called focus-point supersymmetry scenario \cite{focus}.

\vspace{0.2cm}

Obviously, given the above results, 
$\sigma_{\tilde{\chi}_1^0-p}\lsim 3\times 10^{-8}$ pb,
in the mSUGRA scenario with a GUT scale 
more sensitive detectors
producing further data 
are needed.
As discussed in the Introduction,
many dark matter detectors are being projected.
Particularly interesting is the case of GENIUS,
where values of the cross section as low as 
$\approx 10^{-9}$ pb will be accessible,
although this might not be sufficient depending on the values
of the parameters (see Fig.~\ref{cross_scale122}).

\subsection{mSUGRA scenario with an intermediate scale}


The analysis of the cross section
$\sigma_{\tilde{\chi}_1^0-p}$
carried out above in the context of mSUGRA,
was performed assuming the
unification scale
$M_{GUT} \approx 10^{16}$ GeV.
However, there are several interesting phenomenological 
arguments in favour of SUGRA scenarios
with scales 
$M_I\approx 10^{10-14}$ GeV,
such as to explain neutrino masses, the scale
of axion physics, 
and others (see e.g. the discussion in Ref.~\cite{darkcairo} and
references therein).
In addition,
the string scale may be anywhere between the weak and the Planck 
scale, and explicit scenarios with
intermediate scales may arise in the context of 
D-brane constructions from type I strings, 
as we will
discuss in Section~3.
Inspired by all these scenarios, to use the value of the initial scale 
$M_I$ as a free parameter for the running of the soft terms
is particularly interesting.
In fact, it was  
pointed out in Refs.~\cite{muas,Bailin,nosotros} 
that 
$\sigma_{\tilde{\chi}_1^0-p}$
is very sensitive to the variation of $M_I$ 
for the running of the soft terms.
For instance, by taking $M_I=10^{10-12}$ GeV rather than 
$M_{GUT}$,
regions in the parameter space of mSUGRA can be found 
where  $\sigma_{\tilde{\chi}_1^0-p}$ is two orders of magnitude
larger than for $M_{GUT}$ \cite{muas,darkcairo,nosopro,darkufb}.

The fact that smaller scales imply a larger 
$\sigma_{\tilde{\chi}_1^0-p}$  
can be explained with the variation in the value of 
$\mu$ with $M_I$.
One observes that, for $\tan\beta$ fixed, the smaller the initial
scale for the running the smaller the numerator in the
first piece of Eq.~(\ref{electroweak}) becomes. 
This can easily be understood from 
the evolution of $m_{H_u}^2$ with the
scale.
Clearly, when the value of the initial scale is reduced 
the RGE running is shorter and, as a consequence, 
the negative contribution 
$m_{H_u}^2$ to $\mu^2$ in Eq.~(\ref{electroweak}) becomes less important.
Then, 
$|\mu|$ decreases and therefore
the Higgsino composition of the lightest neutralino increases.
Eventually, $|\mu|$ will be of the order of $M_1$, $M_2$
and $\tilde{\chi}_1^0$ will be a mixed Higgsino-gaugino 
state.
In addition, when $|\mu|$ decreases 
$m^2_A=m_{H_d}^2+m_{H_u}^2+2\mu^2$
also decreases. 
As mentioned in the previous Subsection
when talking about increasing $\tan\beta$,
$H$ will decrease and therefore
the scattering channels through Higgs exchange will increase
the cross section.


\begin{figure}
\epsfig{file=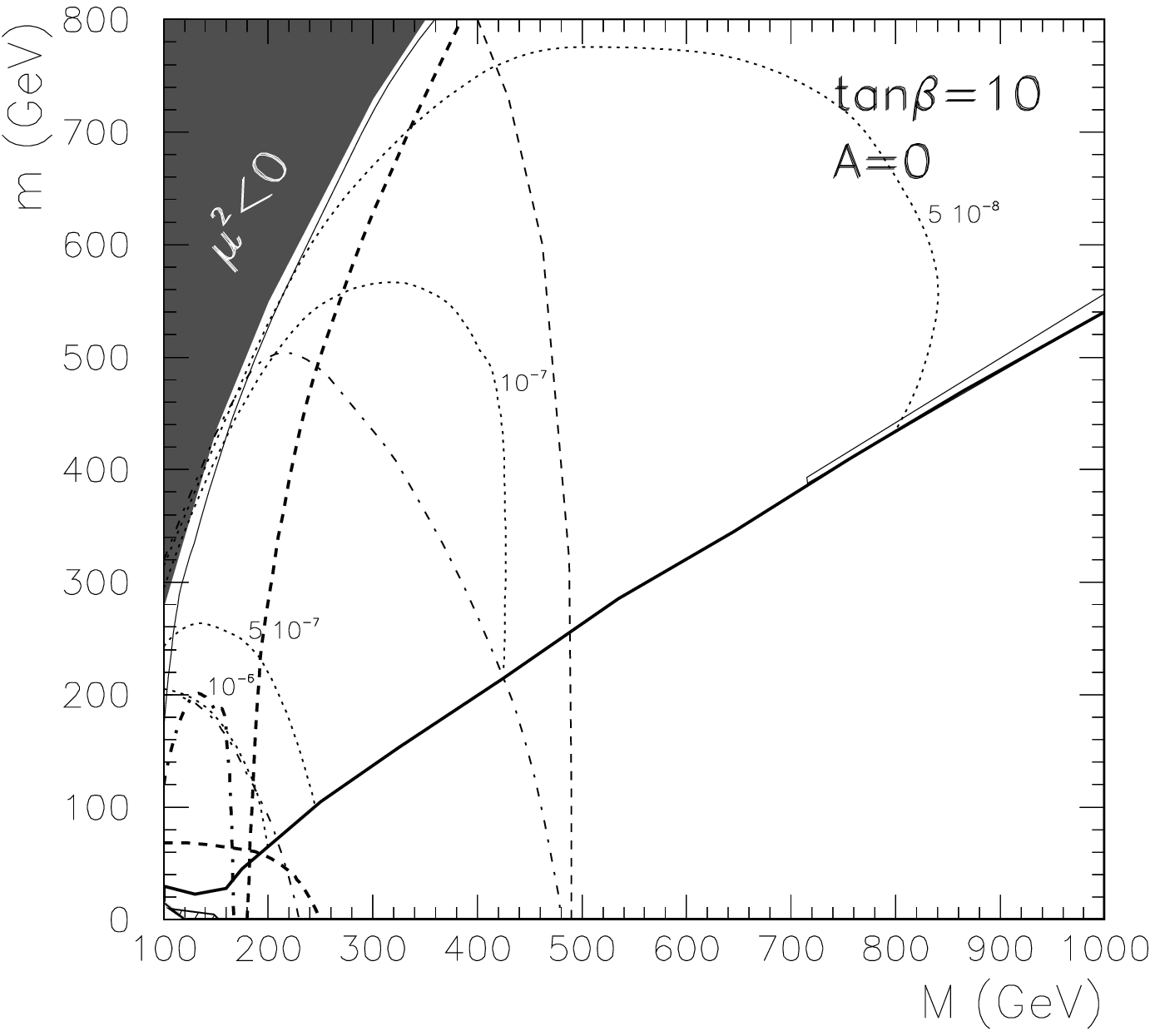,width=7cm}
\epsfig{file=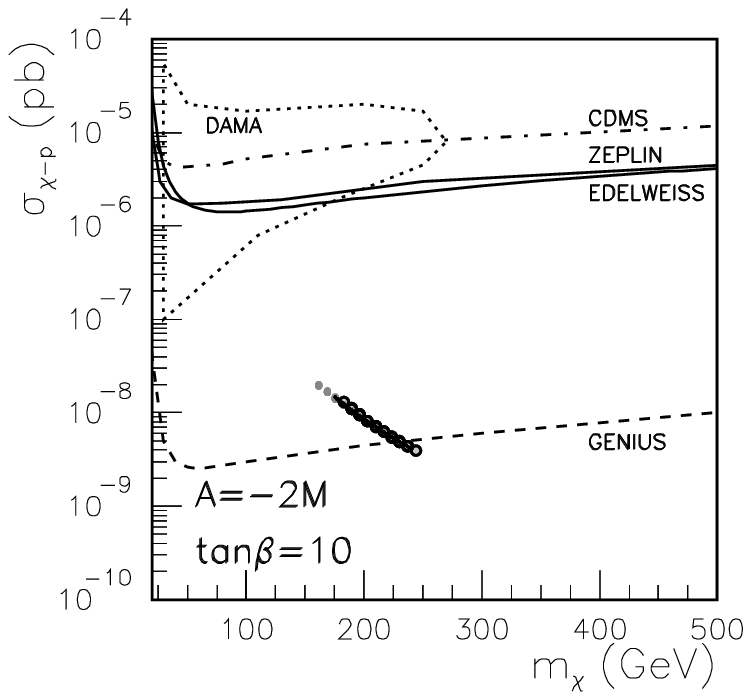,width=7cm}
\caption{{\it Left}: The same as in Fig.~\protect\ref{a2m} but for the 
intermediate scale 
$M_I=10^{11}$. 
The black area is excluded because
$\mu^2$ becomes negative.
The white region at the bottom bounded by a solid line
is excluded because 
$m_{\tilde{\tau}_1}^2$
becomes negative.
{\it Right}: 
The same as in Fig.~\protect\ref{cross_scale122}
but for the intermediate scale
$M_I=10^{11}$ GeV,
with
$\tan\beta=10$ and $A=-2M$.}
\label{scale12_10}
\end{figure}


In the plot on the left frame of
Fig.~\ref{scale12_10} we show the result for $M_I=10^{11}$ GeV, with
$\tan\beta=10$ and $A=0$. 
This case 
can be compared with the one in
Fig.~\ref{a2m}, where $M_{GUT}$ is used. 
Clearly, for the same values of the parameters,
larger cross sections can be obtained with the intermediate scale.
It is worth noticing that even with this moderate value of 
$\tan\beta$, $\tan\beta=10$,
there are regions where the cross section enters in the DAMA 
area, 
$\sigma_{\tilde{\chi}_1^0-p}\approx 10^{-6}$ pb.
However, at the end of the day,
the whole parameter space is forbidden
due to the combination of the Higgs mass bound with
the $g_{\mu}-2$ lower bound. 
Notice also that 
$\Omega_{\tilde{\chi}_1^0}h^2$ is smaller than 0.1 in most of the
parameter space. Only tiny regions bounded by solid lines in the 
figure,
and therefore with 
$0.1\leq \Omega_{\tilde{\chi}_1^0}h^2\leq 0.3$,
can be found.
For other values of $A$, such as
$A=-2M$, there are small regions where the
$m_h$ and $g_{\mu}-2$ bounds
are compatible. Also, 
larger regions with 
$0.1\leq \Omega_{\tilde{\chi}_1^0}h^2\leq 0.3$
are present.
In any case, finally the lower bound for $m_h$ implies
that the allowed cross sections do not enter in the DAMA area.




One also finds that the regions excluded by the UFB-3 constraint 
are much smaller than in those cases where the initial 
scale is the GUT one. 
For example for $A=0$ 
no region is excluded
(see however Fig.~\ref{a2m} for the GUT case). 

In the plot on the right frame of Fig.~\ref{scale12_10} 
we summarize the above results for
$\tan\beta=10$, concerning the
cross section,
showing the values of $\sigma_{\tilde{\chi}_{1}^{0}-p}$
allowed by all experimental constraints 
as a function of the neutralino mass
$m_{\tilde{\chi}_1^0}$, for $A=-2M$.
Only in this case there are dark grey dots corresponding
to points having a relic neutralino density within the
preferred range
$0.1\leq \Omega_{\tilde{\chi}_1^0}h^2\leq 0.3$.
Given the narrow range of these points,
they overlap in the figure with those
excluded by the UFB-3 constraint.

Qualitatively, similar results are obtained for larger values of $\tan\beta$.
For example, for $\tan\beta=35$ 
only for $A=-2M$ we obtain
points allowed by all experimental and astrophysical constraints.
For these, $\sigma_{\tilde{\chi}_{1}^{0}-p}\lsim 10^{-8}$ pb.
Let us finally mention
that in the case
of $\tan\beta=50$,
for $A=-M$ there are points allowed by all experimental and
astrophysical
constraints with
$\sigma_{\tilde{\chi}_{1}^{0}-p}\lsim 10^{-7}$ pb.

\vspace{0.2cm}

Summarizing,
when an intermediate scale is considered in mSUGRA,
although 
the cross section increases significantly
the experimental bounds
impose
$\sigma_{\tilde{\chi}_1^0-p}\lsim 4\times 10^{-7}$ pb.
And, in fact, at the end of the day, the preferred astrophysical range
for the relic neutralino density, 
$0.1\leq\Omega_{\tilde{\chi}_1^0}h^2\leq 0.3$,
imposes 
$\sigma_{\tilde{\chi}_1^0-p}\lsim  10^{-7}$ pb.
Clearly, present experiments are not still sufficient,
and more sensitive detectors
producing further data 
are needed, as in the case of a GUT scale.

\subsection{SUGRA scenario with non-universal gaugino masses 
}

The general situation for 
the soft parameters in SUGRA is to have a
non-universal structure.
For the case of the gaugino masses this is due to the
non-universality of the gauge kinetic functions associated to the
different gauge groups $f_a(h_m)$.
For example, non-universal gaugino masses are obtained if 
the $f_a$ have a different dependence on the hidden sector fields
$h_m$ breaking SUSY.

Analyses of the neutralino-proton cross section using $SU(5)$ GUT relations
for gaugino masses
were carried out in Refs.~\cite{Nath2,Orloff,Roy2}, whereas
in Refs.~\cite{darkcairo,nosopro,Dutta,Birkedal,darkufb} 
generic non-universal soft masses were used. 
It was also realized
that the non-universality in the gaugino masses can increase the cross 
section.

\begin{figure}
\epsfig{file=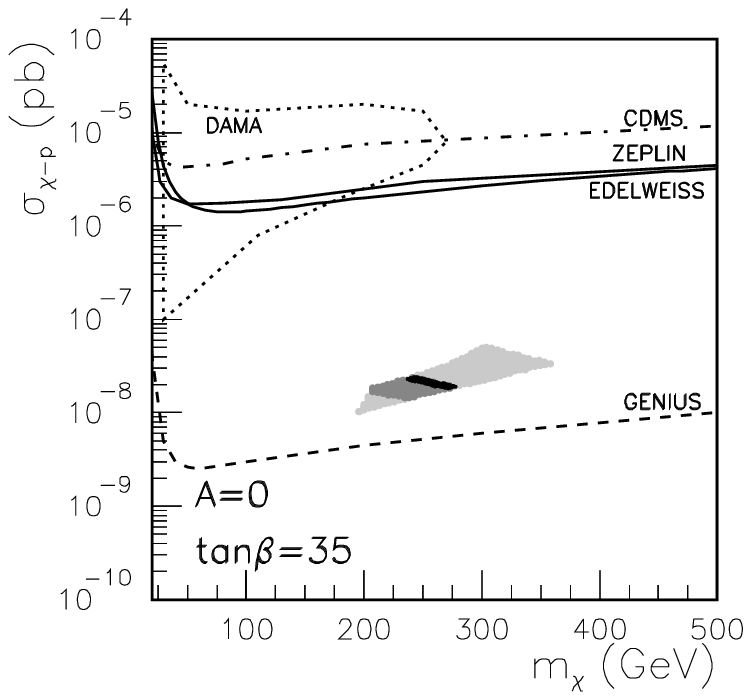,width=7cm}
\hspace*{-0.51cm}\epsfig{file=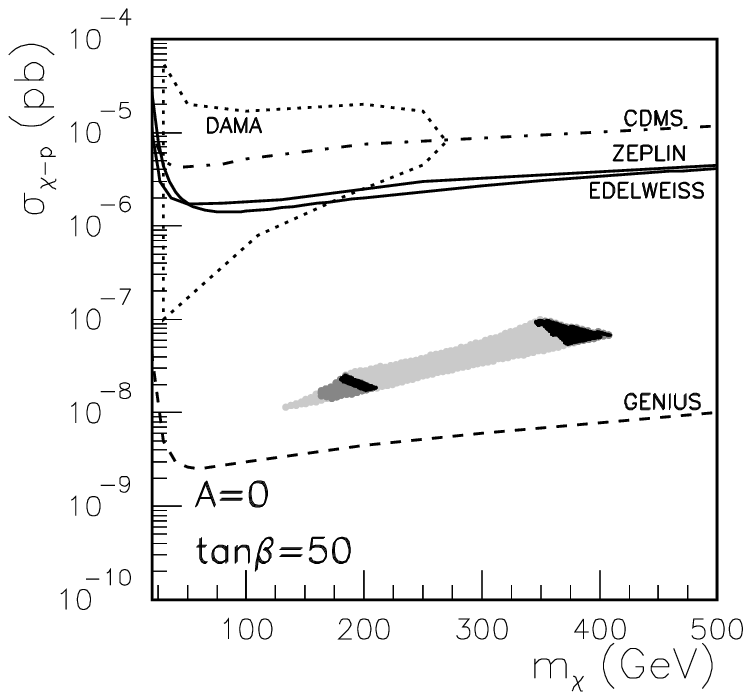,width=7cm}


\caption{The same as in Fig.~\protect\ref{cross_scale122}
but for 
the case discussed in Eq.~(\protect\ref{gauginospara}) with
non-universal soft gaugino masses, 
$\delta'_{1,2}=0, \delta'_3=-0.5$.
}
\label{nuniversal2}
\end{figure}

Let us parameterize this non-universality at $M_{GUT}$
as follows:
\begin{eqnarray}
M_1=M(1+\delta'_{1})\ , \quad M_2=M(1+ \delta'_{2})\ ,
\quad M_3=M(1+ \delta'_{3})
\ ,
\label{gauginospara}
\end{eqnarray}
where $M_{1,2,3}$ are the bino, wino and gluino masses, respectively.
Let us discuss now which values of the parameters are interesting in
order to increase the cross section with respect to the universal
case $\delta'_i=0$. In this sense,
it is worth noticing that
$M_3$ appears in the RGEs of squark masses, so e.g.
their contribution proportional
to the top Yukawa coupling in the RGE of $m_{H_u}^2$ will
do this less negative if $M_3$ is small, and therefore $\mu^2$ will become 
smaller in this case. 
However, small values of $M_3$ also lead to
an important decrease in the Higgs 
mass.
In addition,
$b \to s \gamma$ and $g_{\mu}-2$ constraints are also relevant.

\vspace{0.2cm}

Summarizing, 
although the cross section increases with respect to the universal
case, the present experimental constraints exclude points
entering in the DAMA region.
This is shown in Fig.~\ref{nuniversal2} 
for $\tan\beta=35,50$ and $A=0$, using
$\delta'_{1,2}=0, \delta'_3=-0.5$,
where one can see that
there are 
points allowed by all experimental and astrophysical constraints, but
they
correspond to 
$\sigma_{\tilde{\chi}_1^0-p}\lsim 10^{-7}$ pb.

Finally, let us remark that 
increasing the cross section through values at low energy
of $m_{H_u}^2$ less negatives implies
less important UFB constraints.
Now these are not very relevant.

\subsection{SUGRA scenario with non-universal scalar masses}

As mentioned in the previous subsection, the general situation for 
the soft parameters in SUGRA is to have a
non-universal structure. 
For the case of the observable scalar masses
this is due to the non-universal couplings
in the K\"ahler potential
between the hidden sector fields breaking SUSY and the
observable sector fields.
For example, 
$K=\sum_{\alpha}{\tilde K}_\alpha(h_m,h_m^*) 
\ C_{\alpha} C_{\alpha}^*$, with ${\tilde K}_\alpha$ a function of
the hidden-sector
fields $h_m$, will produce non-universal scalar masses
$m_{\alpha}\neq m_{\beta}$
if ${\tilde K}_\alpha\neq {\tilde K}_\beta$.

Analyses of the dark matter cross section using 
generic non-universal soft masses
were carried out in 
Refs.~\cite{Bottino,arna2,Arnowitt,Santoso,darkcairo,nosopro,Dutta,darkufb},
whereas
in Refs.~\cite{Drees,Nojiri,Arnowitt3,Rosz,Farrill,Profumo}
$SU(5)$ or 
$SO(10)$ GUT relations
were used. 
In the light of the recent experimental results,
one important consequence of the non-universality is that
the cross 
section can be increased in some regions of the parameter 
space 

Let us then parameterized
the non-universality in the Higgs sector as
follows:
\begin{eqnarray}
m_{H_{d}}^2=m^{2}(1+\delta_{1})\ , \quad m_{H_{u}}^{\ 2}=m^{2}
(1+ \delta_{2})\ .
\label{Higgsespara}
\end{eqnarray}
In fact the non-universalities in this sector give the most important
effect, and including the ones in the sfermion sector the cross
section only increases slightly. Thus in what follows we will only
consider
$\delta_{1,2}\neq 0$.

As discussed for intermediate scales in Subsection~2.2,
an important factor in order to increase the cross section 
consists in reducing the value of $|\mu|$.
This value is determined by condition 
$\mu^2\approx -m_{H_u}^2-\frac{1}{2} M_Z^2$
and can
be significantly reduced for some choices of the $\delta$'s. We can have
a qualitative
understanding of the effects of the $\delta$'s on $\mu$ from
the following.
When $m_{H_u}^2$ at $M_{GUT}$ increases,
 i.e. choosing $\delta_2 > 0$,
its negative low-energy contribution to the above condition
becomes less important. 


In addition,
there is another relevant 
way of increasing the cross section using the non-universalities
of the Higgs sector. Note that 
decreasing 
$m_{H_d}^2$, i.e. choosing $\delta_1 < 0$,
leads to a decrease in 
$m^2_A=m_{H_d}^2+m_{H_u}^2+2\mu^2$,
and therefore in the mass of the heaviest 
Higgs $H$.
This produces naturally an increase in the
cross section\footnote{This effect might also be important when
non-universal gaugino masses are taken into account.
The contribution of $M_3$ proportional
to the bottom Yukawa coupling in the RGE of $m_{H_d}^2$ will
do this smaller if $M_3$ is large \cite{prepa}.}.

Thus we will see that,
unlike the previous scenarios, 
with non-universalities in the scalar sector is possible
to obtain large
values of the cross section, and even some points
enter in the DAMA area fulfilling
all constraints. 
Let us analyze three representative cases with \cite{darkufb}
\begin{eqnarray}
a)\,\, \delta_{1}&=&0\ \,\,\,\,\,\,\,\,,\,\,\,\, \delta_2\ =\ 1\ ,
\nonumber\\
b)\,\, \delta_{1}&=&-1\ \,\,\,\, ,\,\,\,\, \delta_2\ =\ 0\ ,
\nonumber\\
c)\,\, \delta_{1}&=&-1\ \,\,\,\, ,\,\,\, \delta_2\ =\ 1 
\ .
\label{3cases}
\end{eqnarray}

Clearly, the above discussion about decreasing $\mu^2$ applies
well to case {\it a)}, where the variation in
$m_{H_u}^2$ through $\delta_2$ is relevant.
For $\tan\beta=35$,
although the cross section increases with respect to the universal
case, and is generically above the GENIUS lower limit,
the present experimental constraints exclude points
entering in the DAMA area. 
We need to go to $\tan\beta=50$ in order to find points
entering in the DAMA area, and 
fulfilling the astrophysical bounds.

\begin{figure}
\epsfig{file=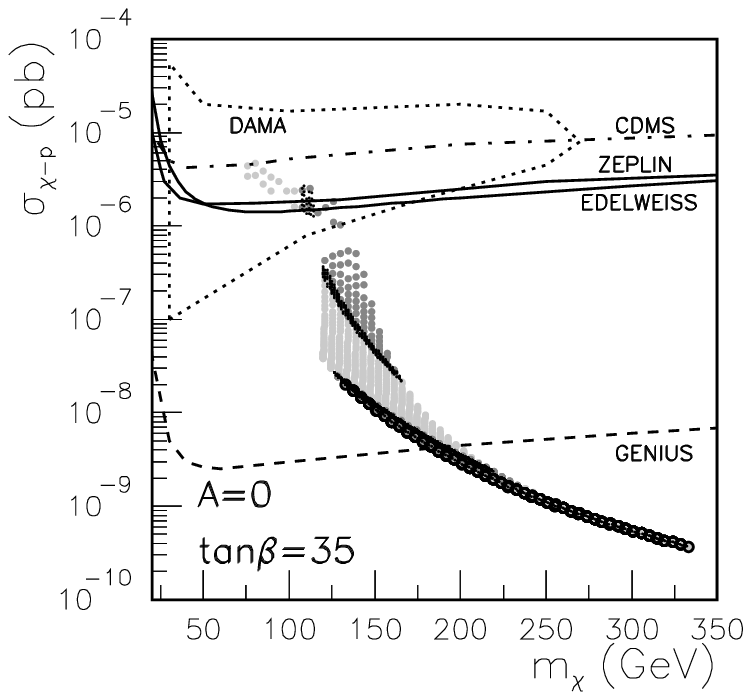,width=7cm}
%
\hspace*{-0.51cm}\epsfig{file=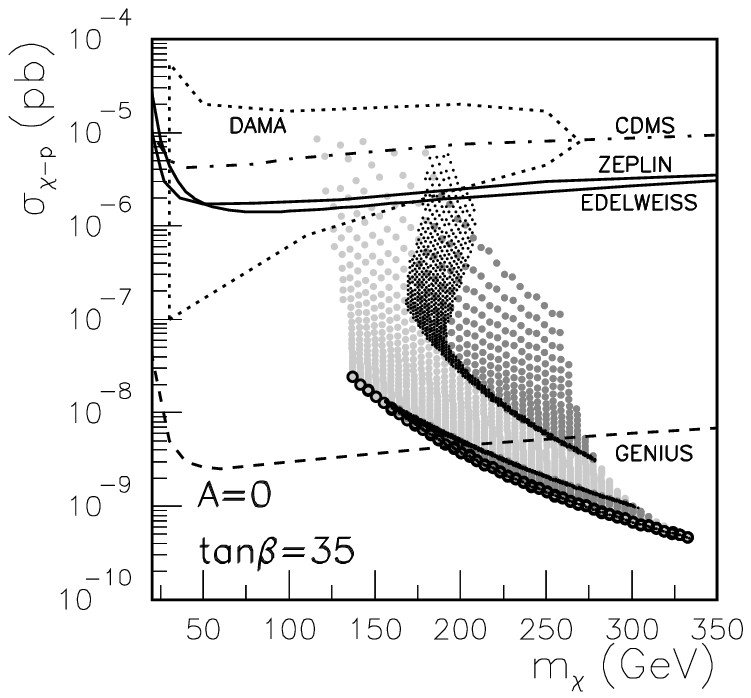,width=7cm}
\caption{The same as in Fig.~\protect\ref{cross_scale122} but for 
the non-universal cases {\it b)} $\delta_1=-1$, $\delta_2=0$ (left),
and {\it c)} $\delta_1=-1$, $\delta_2=1$ (right),
discussed in Eq.~(\protect\ref{3cases}), with $\tan\beta=35$ and $A=0$.}
\label{anusec}
\end{figure}

For case {\it b)} the cross section increases also substantially
with respect to the universal case.
Now $\delta_2$ is taken vanishing and therefore the value 
of $\mu$ is essentially not modified with respect to the universal
case.
However, taking $\delta_1=-1$
produces an increase in the
cross section through 
the decrease in $m_A^2$,
as discussed previously. 
As
shown explicitly in 
in the plot on the left frame of
Fig.~\ref{anusec}, for
$\tan\beta=35$ and $A=0$,
there are points in the DAMA region. 
For $\tan\beta=50$, similarly to case {\it a)}, there are points
entering in the DAMA 
area, and part of them
fulfil the astrophysical bounds. 


Finally, given the above situation concerning the enhancement
of the neutralino-proton cross section for {\it a)} and 
{\it b)}, it is clear that the combination of both cases
might be interesting.
This is carried out
in case {\it c)} where we take $\delta_1=-1$ and 
$\delta_2=1$. As shown 
in the plot on the right frame of
Fig.~\ref{anusec}, cross sections 
as large as $\sigma_{\tilde{\chi}_1^0-p}\gsim 10^{-6}$ pb, entering
in DAMA
and fulfilling all experimental and astrophysical bounds, 
can be obtained for $\tan\beta=35$ and $A=0$.
This is also the case for $\tan\beta=50$. 

\vspace{0.2cm}

In summary,
when
non-universal scalars are allowed in SUGRA, for 
some special choices of the non-universality,  
the cross section can 
be increased a lot with respect to the 
universal scenario.
It is even possible, for some particular values of the
parameters, to find points allowed by all experimental and
astrophysical 
constraints with
$\sigma_{\tilde{\chi}_1^0-p}\approx 10^{-6}$ pb, and therefore
inside the DAMA area\footnote{
This is similar to what occurs in the so-called
effMSSM scenario, where the parameters are
defined directly at the electroweak scale 
\cite{Gondolo}-\cite{Austri}.}. 
Note however that these points would be basically excluded by
the other underground experiments.
In any case, the interesting result is that
large regions accessible for future experiments are present.

\begin{figure}[t]
\begin{center}
\epsfig{file=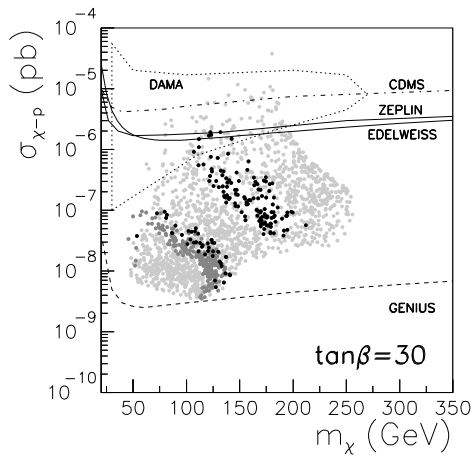,width=7cm}
\end{center}
\caption{
The same as in Fig.~\protect\ref{cross_scale122}
but for a D-brane scenario with string scale 
$M_I=10^{12}$ GeV, and
$\tan\beta=30$.
\label{preliminar}}
\end{figure}

\section{Superstring scenarios}

D-brane constructions are explicit scenarios where two of the interesting
situations studied in the previous Section,
non-universality and intermediate scales, may occur.
Concerning the latter, it was recently 
realized that 
the string scale may be anywhere between the weak and the Plank 
scale (see e.g. the discussion in Ref.~\cite{nosotros} and
references therein),
and in fact 
scenarios with the gauge group and particle content of the
SUSY standard model lead naturally to intermediate values for the
string
scale, in order to reproduce the value of gauge couplings
deduced from experiments \cite{nosotros}. In addition, the soft terms 
turn out to be generically non universal.

Due to these results, in principle, 
large cross sections
might be obtained.
A recent re-analysis allows us to think that this is the case. Very
preliminary results are shown in Fig.~\ref{preliminar},
where one can see that even points entering in DAMA are allowed \cite{prepa}.

\section{Conclusions}

We have carried out a theoretical analysis of the possibility
of detecting dark matter directly in current and projected
experiments. In particular, we have studied the value of the 
neutralino-nucleon
cross section in several supergravity and superstring scenarios.
In addition to the usual experimental and astrophysical constraints
we have imposed on the parameter space 
the absence of dangerous charge
and colour breaking minima.

In the usual mSUGRA scenario, where the soft terms
are assumed to be universal, and the GUT scale is considered, 
the cross section is
constrained to be 
$\sigma_{\tilde{\chi}_1^0-p}\lsim 3\times 10^{-8}$ pb.
Obviously, in this case, present experiments are not sufficient and
more sensitive detectors
producing further data 
are needed.
A similar conclusion is obtained when an intermediate scale is considered.
Although 
the cross section increases significantly,
the experimental bounds
impose
$\sigma_{\tilde{\chi}_1^0-p}\lsim 4\times 10^{-7}$ pb.
And, in fact, at the end of the day, the preferred astrophysical range
for the relic neutralino density, 
$0.1\leq\Omega_{\tilde{\chi}_1^0}h^2\leq 0.3$,
imposes 
$\sigma_{\tilde{\chi}_1^0-p}\lsim  10^{-7}$ pb.
Still present experiments are not sufficient.

When
non-universal scalars are allowed in SUGRA, for 
some special choices of the non-universality,  
the cross section can 
be increased a lot with respect to the 
universal scenario.
It is even possible, for some particular values of the
parameters, to find points allowed by all experimental and
astrophysical 
constraints with
$\sigma_{\tilde{\chi}_1^0-p}\approx 10^{-6}$ pb, and therefore
inside the DAMA area. 
For non-universal gauginos,
although the cross section increases, the experimental bounds
exclude this possibility. 

On the other hand, 
the low-energy limit of superstring theory is 
SUGRA, and therefore the neutralino will also be a candidate for
dark matter in these scenarios.
In this context we have studied
D-branes configurations from type I string, where intermediate
scales and non-universal soft terms
arise naturally.
A preliminary analysis allows us to think
that some regions of the parameter space may have large values
of the cross section, even compatible 
with DAMA.

Finally,
it is worth recalling that although the DAMA result is controversial
because the negative search result obtained by other recent
experiments,
our main point is that large regions of the parameter space accessible
for  experiments are present in several SUSY scenarios.

\vspace{0.5cm}

\noindent {\bf Acknowledgements}

\noindent
We thank D.G. Cerde\~no, 
E. Gabrielli and M.E. Gomez 
as co-authors of the works reported
in this talk.


\end{document}